\documentclass[preprint]{article}
\usepackage{authblk}
\usepackage{graphicx}
\usepackage{dcolumn}
\usepackage{bm}
\usepackage{amsmath}
\usepackage{amssymb}
\usepackage{multirow}
\usepackage{caption}
\usepackage{subcaption}
\usepackage{epstopdf}
\usepackage{hyperref}
\usepackage{cite}
\begin{document}
\title{\bf Observational signatures of scalar field dynamics in modified $f(Q, L_m)$ gravity}
\author[]{Amit Samaddar\thanks{samaddaramit4@gmail.com}}
\author[]{S. Surendra Singh\thanks{ssuren.mu@gmail.com}}
\affil[]{Department of Mathematics, National Institute of Technology Manipur, Imphal-795004,India.}

\maketitle
 
\textbf{Abstract}: We investigate the cosmological implications of a tanh-parametrized scalar field model in the framework of modified $f(Q, L_{m})$ gravity by adopting the form $f(Q, L_{m})=\beta Q+\delta L_{m}$ along with a scalar field energy density $\rho_\phi = \rho_{c0} \tanh(A + Bz)$. Using MCMC methods and combining $31$ cosmic chronometer data points, $15$ BAO, DESI DR2 BAO and $1701$ Pantheon+ samples, we constrain the model parameters and obtain $H_{0}=74.284^{+4.155}_{-4.275}$, $\Omega_{m0}=0.326^{+0.093}_{-0.072}$ and $B=-0.001^{+0.030}_{-0.030}$. The model predicts a transition redshift $z_{tr}=0.5914$ and a present deceleration parameter $q_0=-0.5167$, consistent with a Universe transitioning from deceleration to acceleration. We further analyze the evolution of the EoS parameter, density components and statefinder diagnostics in which all parameters show asymptotic convergence to a de Sitter phase. Additionally, we study black hole mass accretion, showing its dependence on the scalar field dynamics. This work highlights the compatibility of tanh-scalar field forms with $f(Q, L_m)$ gravity in describing cosmic acceleration and gravitational phenomena.


\section{Introduction}\label{sec1}
\hspace{0.5cm} The immense and intricate nature of the Universe has long fueled human curiosity and scientific inquiry. Over the past few decades, breakthroughs in observational astronomy have shown that the Universe is not just expanding, but accelerating in its expansion—a remarkable discovery initially made through supernova observations in the late $1990$s \cite{Riess98,Per99}. Such cosmic acceleration implies the dominance of an unknown entity—dark energy (DE)—which is believed to constitute nearly $70\%$ of the Universe’s energy density. Although the $\Lambda$CDM model successfully accounts for a wide range of cosmological data, it falls short in addressing some fundamental issues. Notably, the cosmological constant problem and the coincidence problem remain unresolved—raising critical questions about why the vacuum energy is extremely small and why its effects have become dominant only in the current phase of the Universe's evolution \cite{WMAP13, Pee03}. The persistent mismatch between Hubble constant measurements derived from early-Universe observations and those based on late-Universe data—commonly referred to as the Hubble tension—highlights the growing demand for novel theoretical explanations \cite{Steven89}.

The foundation of these cosmological issues may lie in our understanding of gravity. Although General Relativity (GR) has proven immensely successful, it may not be fully adequate at describing the Universe on the largest scales. Within GR, gravity arises from the curvature of spacetime, yet the observed accelerated expansion necessitates the inclusion of theoretical elements with unclear physical nature. To address this, modified gravity theories have emerged as promising alternatives, aiming to reproduce GR’s predictions in local settings while providing new mechanisms to account for late-time cosmic acceleration \cite{Clifton12}. A broad spectrum of modified gravity models has been proposed, including $f(R)$, $f(G)$ and $f(R,L_{m})$ theories, where the gravitational action is extended through functions involving curvature invariants and matter contributions. These frameworks introduce curvature-matter couplings, giving rise to a variety of gravitational dynamics \cite{Harko10, Wang12, Nojiri11}. In recent years, interest has shifted toward alternative geometrical formulations that rely on torsion or non-metricity instead of curvature. Among them, the symmetric teleparallel formulation—based on the non-metricity scalar $Q$—serves as the basis for $f(Q)$ gravity, where gravitational effects arise from variations in the metric under parallel transport rather than from spacetime curvature \cite{Jimenez18, Kh21}.

An especially intriguing development within this framework is the $f(Q,L_{m})$ gravity theory, where the gravitational action incorporates both the non-metricity scalar $Q$ and the matter Lagrangian $L_{m}$ \cite{Hazarika24}. This formulation establishes a direct coupling between matter and geometry, leading to significant modifications in the structure of the gravitational field equations. Such an interaction gives rise to complex cosmological behavior, including potential effective interactions between dark matter and dark energy, along with enhanced consistency with observational data. Moreover, the model’s adaptability enables it to address a wide range of phenomena, from the early Universe—such as inflation and baryogenesis—to the late-time accelerated expansion \cite{K2024, Y2024, Y2025, A2025}. Consequently, $f(Q,L_{m})$ gravity emerges as a promising and dynamic avenue in the pursuit of a deeper understanding of the cosmos.

Scalar fields have been instrumental in shaping our understanding of the Universe, especially in the context of DE and its evolving nature. One of the most widely studied models is quintessence—a scalar field alternative to the cosmological constant that permits a time-dependent energy density and an equation of state parameter $-1<\omega<-0.33$. Unlike the static vacuum energy in the $\Lambda$CDM model, quintessence introduces a dynamic mechanism for cosmic acceleration, governed by the balance between its kinetic and potential components. When the potential energy dominates over the kinetic term $(\dot{\phi}^{2}<<V(\phi))$, the field behaves like a repulsive gravitational force, accelerating the Universe’s expansion. However, if kinetic energy becomes dominant, the acceleration can diminish or even cease. This adaptability enables scalar field models to naturally account for the transition from an early decelerating phase to the current accelerated expansion \cite{SM1998,Cope06,Martin08,Amend00,Singh23}. Beyond the standard quintessence model, numerous alternative scalar field theories have been proposed to shed light on the dark energy puzzle and the Universe's accelerating expansion. One such variant is the phantom field, which features an equation of state parameter $\omega<-1$, which describes a more exotic form of dark energy that could lead to a super-accelerated expansion and potentially culminate in a cosmic singularity, like the so-called ``Big Rip" \cite{Tsu05,RR03}. Another significant development is the class of k-essence models, which generalize the scalar field Lagrangian by incorporating non-standard kinetic terms. These models generate a broad array of dynamical behaviors, making them capable of addressing both inflationary dynamics in the early Universe and the current phase of acceleration \cite{Chiba00,CA00}. The Chaplygin gas model offers a distinct approach by attempting to unify dark matter and dark energy into a single fluid governed by a unique equation of state. This model transitions from a matter-like phase in the early Universe to a dark energy-dominated regime at late times, providing a unified narrative for cosmic evolution \cite{AY01,MC02,AY001}. Tachyon fields, inspired by string theory, represent another compelling class. Under slow-roll conditions, they can replicate dark energy behavior, making them strong contenders in the search for a dynamical DE model \cite{Bagla03}. Further extending the landscape, models incorporating fermionic fields and their interactions with scalar components—such as g-essence—have been formulated to introduce more complex dynamics and potential observational signatures \cite{NN19,KK12}. These hybrid models expand the theoretical framework available for interpreting cosmic acceleration and pave the way for more comprehensive cosmological theories. Taken together, these diverse scalar field scenarios emphasize the rich variety of approaches available for understanding the accelerating Universe and demonstrate the value of exploring beyond conventional models.

In recent years, scalar field models have increasingly been explored using data-driven parametrizations that aim to reconstruct the nature of dark energy from observational insights. Notably, \cite{JK20} proposed the Exponential Decay Scalar Field Dynamics (EDSFD) model to analyze scalar field evolution within a homogeneous and isotropic FLRW universe, presenting both analytical and numerical insights into its impact on cosmic dynamics. Their analysis also assessed the stability of various dynamical regimes. Complementary work by \cite{Pacif21} employed diagnostic tools such as the statefinder parameters to differentiate among dark energy models, confirming the viability of scalar field scenarios with observational probes like supernovae, baryon acoustic oscillations (BAO), and cosmic microwave background (CMB) data. The intersection of scalar field dynamics with modified gravity theories has also gained considerable attention. For example, \cite{Kar22} examined how Dirac–Born–Infeld (DBI) scalar fields coupled with $f(Q)$ gravity influence astrophysical mass accretion phenomena. Other approaches have incorporated scalar fields into the Barrow holographic dark energy model within the $f(R,T)$ gravity framework, adding further depth to the theoretical landscape \cite{Sharma22}. Further, \cite{Wang23} examined a logarithmic parametrization of scalar field energy density, offering a novel approach to model the transition from deceleration to acceleration in the cosmic expansion history.

Motivated by this growing body of research, our study adopts a parametric reconstruction strategy to model the scalar field evolution without committing to a specific potential from the outset. Instead, we express the scalar energy density $\rho_{\phi}$ as a redshift-dependent function, choosing a tanh form to reflect the transition from early deceleration to late-time acceleration. This approach enables a phenomenologically rich yet physically motivated description of cosmic evolution that remains directly testable. By comparing the model’s predictions with observational data—such as cosmic chronometers, BAO (including DESI DR2) and Type Ia supernovae—we constrain its parameters and evaluate its consistency with the Universe’s expansion history.

To further validate the viability of our scalar field parametrization and its consistency with the observed expansion history of the Universe, we adopt an observational data-driven approach. While the theoretical framework provides the foundation, empirical constraints are essential to distinguish among cosmological models and reduce degeneracy in parameter space. In this work, we utilize a combination of recent cosmological datasets—specifically CC, BAO (including DESI DR2) and Type Ia supernovae (SN Ia)—which probe different epochs and aspects of the Universe's evolution. These complementary datasets enhance the robustness of our analysis by covering both the early and late Universe, allowing us to tightly constrain the model parameters and assess the physical plausibility of the proposed scalar field dynamics. This approach not only enables a rigorous examination of the model’s predictions but also facilitates comparison with standard and competing cosmological models.

The organization of this manuscript is as follows. In section \ref{sec2}, we provide a concise overview of the field equations in the framework of $f(Q,L_{m})$ gravity. Section \ref{sec3} introduces a specific form of the gravitational Lagrangian and solves the resulting modified field equations. In section \ref{sec4}, we introduce a redshift-dependent parametrization for the scalar field energy density and derive the corresponding expression for the Hubble parameter. Section \ref{sec5} focuses on the observational constraints to fit the model parameters. In section \ref{sec6}, we examine the evolution of fundamental cosmological quantities, providing a detailed assessment of the model's dynamical characteristics. Section \ref{sec7} is devoted to computing the cosmic age and examining its compatibility with observational estimates. In section \ref{sec8}, we explore the statefinder diagnostics to distinguish our model from other dark energy scenarios. The implications of our scalar field parametrization on black hole mass accretion are addressed in section \ref{sec9}. Finally, section \ref{sec10} summarizes our findings and discusses possible future extensions of this work.
\section{Mathematical structure of $f(Q,L_{m})$ gravity}\label{sec2}
\hspace{0.5cm} This research delves into a novel formulation of symmetric teleparallel gravity, where the gravitational action is constructed from
\begin{equation}\label{1}
S=\int f(Q,L_{m})\sqrt{-g}d^{4}x,
\end{equation}
with $\sqrt{-g}$ denotes the metric tensor's determinant, which preserves the general covariance of the theory. The $f(Q,L_{m})$ function is defined to depend freely on non-metricity scalar $Q$ and the matter-Lagrangian $L_{m}$. This leads to a more comprehensive theoretical framework, going beyond the traditional symmetric teleparallel and general relativity descriptions. According to \cite{Jimenez18}, the non-metricity scalar $Q$ is specified as
\begin{equation}\label{2}
Q=-g^{\alpha\beta}\big(L^{\nu}{}_{\mu\alpha}L^{\mu}{}_{\beta\nu}-L^{\nu}{}_{\mu\nu}L^{\mu}{}_{\alpha\beta}\big),
\end{equation}
Here, $L^{\nu}{}_{\mu\eta}$ representing the disformation tensor, expressed explicitly as
\begin{equation}\label{3}
L^{\nu}{}_{\mu\eta}=\frac{g^{\nu\gamma}}{2}\big[Q_{\eta\mu\gamma}+Q_{\mu\gamma\eta}-Q_{\gamma\mu\eta}\big],
\end{equation} 
The non-metricity tensor measures the deviation from metric compatibility, i.e., it characterizes how the metric tensor fails to remain invariant under covariant differentiation. It is defined as
\begin{equation}\label{4}
Q_{\eta\alpha\beta}=-\nabla_{\eta}g_{\alpha\beta}=-\partial_{\eta}g_{\alpha\beta}+g_{\beta\delta}\breve{\Gamma}^{\delta}{}_{\alpha\eta}+g{\delta\alpha}\breve{\Gamma}^{\delta}{}_{\beta\eta},
\end{equation}
Here, $\breve{\Gamma}^{\delta}{}_{\alpha\beta}$ refers to the connection associated with the symmetric teleparallel geometric framework. The non-metricity tensor admits two important contractions, which are defined as follows:
\begin{equation}\label{5}
Q_{\nu}=g^{\alpha\beta}Q_{\nu\alpha\beta}, \hspace{0.5cm} \breve{Q}_{\nu}=g^{\alpha\beta}Q_{\alpha\nu\beta},
\end{equation}
These trace components play a central role in the formulation of the theory. To assist in deriving the field equations, we define a tensor often referred to as the superpotential or the conjugate of non-metricity, denoted by
\begin{equation}\label{6}
P^{\nu}{}_{\alpha\beta}\equiv\frac{1}{4}\left[-Q^{\nu}{}_{\alpha\beta}+2 Q^{\;\;\nu}_{(\alpha\;\;\beta)}+Q^{\nu}g_{\alpha\beta}-\tilde{Q}^{\nu}g_{\alpha\beta}-\delta^{\nu}_{\;(\alpha}Q_{\beta)} \right].
\end{equation}
Alternatively, this tensor can be recast in terms of the disformation tensor $L^{\nu}{}_{\alpha\beta}$ as
\begin{equation}\label{7}
P^{\nu}{}_{\alpha\beta}=-\frac{1}{2}L^{\nu}{}_{\alpha\beta}+\frac{1}{4}\left(Q^{\nu}-\tilde{Q}^{\nu}\right)g_{\alpha\beta}-\frac{1}{4}\delta^{\nu}_{\;(\alpha}Q_{\beta)}.
\end{equation}
The superpotential serves as a key geometric object that encodes the non-metric aspects of the connection and appears prominently in the variation of the action. The non-metricity scalar $Q$ can be expressed through its contraction with the superpotential tensor $P^{\nu}{}_{\alpha\beta}$ as follows:
\begin{equation}\label{8}
Q=-Q_{\nu\alpha\beta} P^{\nu\alpha\beta}=-\frac{1}{4}\left[-Q_{\nu\beta\epsilon}Q^{\nu\beta\epsilon}+2Q_{\nu\beta\epsilon}Q^{\epsilon\nu\beta}-2Q_{\epsilon}\tilde{Q}{\epsilon}+Q_{\epsilon}Q^{\epsilon}\right].
\end{equation}
This expression reflects how different contractions and traces of the non-metricity tensor contribute to the scalar invariant $Q$, which governs the geometric sector of the theory. To derive the gravitational field equations, one performs a variation of the action (\ref{1}) with respect to the metric tensor $g_{\alpha\beta}$. This leads to the following form of the field equations:
\begin{equation}\label{9}
\frac{2}{\sqrt{-g}}\nabla_{\mu}\left(f_{Q}\sqrt{-g}P^{\nu}{}_{\alpha\beta}\right)+f_{Q}\left(P_{\alpha\mu\nu}Q_{\beta}^{\;\;\mu\nu}-2Q^{\mu\nu}_{\;\alpha}P_{\mu\nu\beta}\right)+\frac{f}{2} g_{\alpha\beta}=\frac{f_{L_{m}}}{2}\left(g_{\alpha\beta}L_{m}-T_{\alpha\beta}\right).
\end{equation}
These equations describe how geometry, encoded via non-metricity, interacts with matter fields through the function $f(Q,L_{m})$, offering a broad generalization of standard gravitational dynamics. In these equations, the derivatives of the function $f(Q,L_{m})$ are denoted by $f_{Q}=\frac{\partial f}{\partial Q}$ and $f_{L_{m}}=\frac{\partial f}{\partial L_{m}}$. A particularly important case arises when the function splits additively as $f(Q,L_{m})=f(Q)+2L_{m}$. Under this condition, the theory simplifies to standard $f(Q)$ gravity, as shown in \cite{Jimenez18}. The energy–momentum tensor associated with the matter fields is defined through variation of the matter action with respect to the metric and takes the form:
\begin{equation}\label{10}
T_{\alpha\beta}=-\frac{2}{\sqrt{-g}}\frac{\delta(\sqrt{-g}L_{m})}{\delta g^{\alpha\beta}} = g_{\alpha\beta}L_{m}-2\frac{\partial L_{m}}{\partial g^{\alpha\beta}}.
\end{equation}
A further variation of the action with respect to the affine connection leads to an additional dynamical equation:
\begin{equation}\label{11}
\nabla_{\alpha}\nabla_{\beta}\left[4\sqrt{-g}f_{Q}P^{\alpha\beta}_{\;\;\;\mu}+H_{\mu}^{\;\;\alpha\beta}\right]=0,
\end{equation}
where the term $H_{\mu}^{\;\;\alpha\beta}$ denotes the hypermomentum density, which accounts for the dependence of the matter Lagrangian on the connection. This quantity is defined by:
\begin{equation}\label{12}
H_{\mu}^{\;\;\alpha\beta}=\sqrt{-g}f_{L_{m}}\frac{\delta L_{m}}{\delta Y^{\mu}_{\;\;\alpha\beta}},
\end{equation}
with $Y^{\mu}_{\;\;\alpha\beta}$ representing the distortion tensor, encoding how matter couples to the connection beyond the metric. The presence of hypermomentum distinguishes this theory from purely metric ones and highlights its metric-affine character.

One of the defining characteristics of $f(Q,L_{m})$ gravity is the violation of energy-momentum conservation. When taking the covariant derivative of the field equations derived earlier, one obtains the following relation:
\begin{equation}\label{13}
D_{\alpha}T^{\alpha}{}_{\beta}=\frac{1}{f_{L_{m}}}\bigg(\frac{2}{\sqrt{-g}}\nabla_{\mu}\nabla_{\alpha}H^{\;\;\mu\alpha}_{\beta}+\nabla_{\alpha}A^{\alpha}_{\beta}-\nabla_{\alpha}\bigg[\frac{1}{\sqrt{-g}}\nabla_{\mu}
H^{\;\;\mu\alpha}_{\beta}\bigg]\bigg)=B_{\beta}\neq 0.
\end{equation}
This equation clearly signals that the energy-momentum tensor $T^{\alpha}{}_{\beta}$ is not divergenceless, i.e., energy and momentum are not conserved in the usual sense. The term $B_{\beta}$, which measures this non-conservation, emerges from the interaction between the matter fields and the underlying non-metric geometric structure. It depends explicitly on the choice of the function $f(Q,L_{m})$, the specific form of the matter Lagrangian $L_{m}$ and the behavior of other dynamical fields in the theory.

To study the dynamics of the Universe in the context of $f(Q,L_{m})$ gravity, we work within a flat Friedmann–Lema$\hat{i}$tre–Robertson–Walker spacetime. This cosmological model relies on the principles of homogeneity, implying the Universe is evenly structured on large scales, and isotropy, indicating no preferred direction in space. The FLRW metric is well-suited for modeling the expanding Universe under these symmetries. The spacetime geometry is described by the line element
\begin{equation}\label{14}
ds^{2}=-dt^{2}+a^{2}(t)(dx^{2}+dy^{2}+dz^{2}),
\end{equation}
where $a(t)$ denotes the scale factor, which evolves with cosmic time $t$ and characterizes the expansion of the Universe. In this cosmological setting, the non-metricity scalar reduces to a simple form: $Q=6H^{2}$, where the Hubble parameter $H=\frac{\dot{a}}{a}$ quantifies the rate at which the Universe expands. To represent the matter distribution in the Universe, we assume it behaves as a perfect fluid. Its energy-momentum structure is characterized by the tensor: 
\begin{equation}\label{15}
T_{\alpha\beta}=(p+\rho)u_{\alpha}u_{\beta}+pg_{\alpha\beta},
\end{equation}
In this case, $\rho$ represents the energy density of the fluid, $p$ is the isotropic pressure exerted by the fluid and $u_{\alpha}$ denotes the four-velocity field describing the motion of the fluid through spacetime.

Using the FLRW metric from equation (\ref{14}) alongside the perfect fluid form of the energy-momentum tensor given in (\ref{15}), and applying the variational principle to the modified gravitational action, we derive the modified Friedmann equations that describe the evolution of the Universe in $f(Q,L_{m})$ gravity \cite{Hazarika24}. These equations take the form:
\begin{equation}\label{16}
3H^{2}=\frac{1}{4f_{Q}}\bigg[f-f_{L_{m}}(\rho+L_{m})\bigg],
 \end{equation}
 \begin{equation}\label{17}
\dot{H}+3H^{2}+\frac{\dot{f_{Q}}}{f_{Q}}H=\frac{1}{4f_{Q}}\bigg[f+f_{L_{m}}(p-L_{m})\bigg].
 \end{equation}
 \section{Cosmological framework with linear $f(Q,L_{m})$ coupling}\label{sec3}
 \hspace{0.5cm} To explore the cosmological consequences of $f(Q,L_{m})$ gravity, we adopt a simple yet insightful linear form for the function $f$, given by:
 \begin{equation}\label{18}
 f(Q,L_{m})=\beta Q+\delta L_{m},
 \end{equation}
 where the constants $\beta$ and $\delta$ determine the coupling strengths to the non-metricity scalar $Q$ and the matter Lagrangian $L_{m}$, respectively. This linear formulation serves as a natural starting point for analysis due to its tractability and is motivated by analogous constructions in modified gravity theories—such as the models $f(R,T)=R+2\lambda T$ \cite{Harko11}, $f(Q,C)=\gamma Q+\delta C$ \cite{Amits24}, $f(Q,T)=\gamma Q+\delta T$ \cite{Xu20}, $f(Q,L_{m})=\gamma Q+\delta L_{m}$ \cite{Y2025} and $f(Q,L_{m})=-\gamma Q+2L_{m}+\delta$ \cite{Y2024}-which have demonstrated that even minimal coupling terms can produce significant modifications to cosmic dynamics, including the potential for explaining late-time acceleration.
 
 For the special case where $\beta=1$ and $\delta=0$, the theory reduces to standard General Relativity, highlighting how the linear model extends GR in a straightforward yet impactful manner. In line with previous studies such as those in $f(R,L_{m})$ \cite{Ober08}, $f(Q,L_{m})$ \cite{Kavya22} and $f(G)$ \cite{Zhao12}, we adopt the matter Lagrangian as $L_{m}=-\rho$, which is a common and consistent choice in perfect fluid cosmology under the comoving frame. Substituting this linear form into the general field equations simplifies the dynamics considerably. Since the partial derivatives become constants, we have:
 \begin{equation}\label{19}
 f_{Q}=\beta, \hspace{0.5cm} f_{L_{m}}=\delta,
 \end{equation}
 With this, the generalized Friedmann equations reduce to the following modified forms:
  \begin{equation}\label{20}
 H^{2}=-\frac{\delta\rho}{6\beta},
 \end{equation}
 \begin{equation}\label{21}
 2\dot{H}+3H^{2}=\frac{\delta p}{2\beta}.
 \end{equation}
\section{Scalar field dynamics and its cosmological role}\label{sec4}
\hspace{0.5cm} To deepen our understanding of the cosmic evolution in the $f(Q,L_{m})$ gravity framework, we extend the model by incorporating a scalar field $\phi$, which has long been considered a compelling candidate for describing early-universe inflation, dark energy (DE) and phenomena such as the Higgs mechanism. The scalar field’s behavior is governed by its action, which takes the general form:
\begin{equation}\label{22}
S_{\phi}=\int\bigg[\frac{1}{2}\partial^{\alpha}\phi\partial_{\beta}\phi-V(\phi)\bigg]\sqrt{-g}d^{4}x,
\end{equation}
where $V(\phi)$ is the self-interaction potential and $\partial^{\alpha}\phi$ captures the kinetic contributions of the field. Varying this action with respect to $\phi$ yields the well-known Klein–Gordon equation in the context of an expanding Universe:
\begin{equation}\label{23}
\ddot{\phi}+3H\dot{\phi}+\frac{dV}{d\phi}=0,
\end{equation}
where the dot denotes differentiation with respect to cosmic time $t$. The energy-momentum tensor associated with the scalar field, responsible for its gravitational effects, can be expressed as:
\begin{equation}\label{24}
T^{\phi}_{\alpha\beta}=(\rho_{\phi}+p_{\phi})u_{\alpha}u_{\beta}-p_{\phi}g_{\alpha\beta},
\end{equation}
where the energy density and pressure contributions from the scalar field are:
\begin{equation}\label{25}
\rho_{\phi}=\frac{\dot{\phi}^{2}}{2}+V(\phi), \hspace{0.5cm} p_{\phi}=\frac{\dot{\phi}^{2}}{2}-V(\phi),
\end{equation}
These two terms represent the kinetic and potential energy components of the field and their interplay drives the acceleration or deceleration of the cosmic scale factor. In our cosmological setup, we assume the Universe is composed of two major components: $(a)$ pressureless matter (e.g., cold dark matter) with $p_{m}=0$ and $(b)$ a time-evolving scalar field $\phi$, acting as a source of dark energy. Hence, the Universe's total energy density and pressure are determined to be:
\begin{equation}\label{26}
\rho=\rho_{m}+\rho_{\phi}, \hspace{0.5cm} p=p_{m}+p_{\phi},
\end{equation}
The conservation law governing the cosmic fluid's total behavior reads:
\begin{equation}\label{27}
\dot{\rho}_{m}+\dot{\rho}_{\phi}+3H(\rho_{m}+\rho_{\phi}+p_{\phi})=0,
\end{equation}
Assuming matter and scalar field components conserve independently without interaction, the conservation equations are separated into:
\begin{equation}\label{28}
\dot{\rho}_{m}+3H\rho_{m}=0,
\end{equation}
and
\begin{equation}\label{29}
\dot{\rho}_{\phi}+3H(\rho_{\phi}+p_{\phi})=0,
\end{equation}
By integrating the conservation equation (\ref{28}) for pressureless matter, we derive its evolution in terms of redshift as:
\begin{equation}\label{30}
\rho_{m}=\rho_{m0}(1+z)^{3},
\end{equation}
where $\rho_{m0}$ is the present-day matter energy density. The scalar field equation gives an evolution equation for $\rho_{\phi}$:
\begin{equation}\label{31}
\dot{\rho}_{\phi}+3H\rho_{\phi}(1+\omega_{\phi})=0\;\; \Rightarrow\;\; \omega_{\phi}=-1-\frac{\dot{\rho}_{\phi}}{3H\rho_{\phi}}.
\end{equation}
Here, $\omega_{\phi}$ represents the scalar field’s equation of state parameter.

We examine the role of a dynamic scalar field as a dark energy candidate in our cosmological framework by using a phenomenological representation of the scalar field energy density in terms of redshift, given by:
\begin{equation}\label{32}
\rho_{\phi}=\rho_{c0}tanh(A+Bz),
\end{equation}
where $\rho_{c0}$ is the critical density at present and $A$ and $B$ are the model parameters. This functional form is not derived from a specific scalar potential but is chosen based on its ability to effectively capture a smooth transition in the cosmic energy budget across time. The motivation behind this assumption is threefold. First, observational data indicate that dark energy is subdominant at high redshifts (early times) and becomes significant only in the more recent cosmological epochs. The hyperbolic tangent function reflects this behavior naturally: it evolves slowly and remains suppressed at large $z$, mimicking a negligible dark energy contribution in the early Universe, and asymptotically saturates to a constant at low redshifts, consistent with the current dark energy-dominated phase. This smooth interpolation makes it particularly suitable for modeling evolving dark energy without introducing singularities or abrupt transitions. Second, from a theoretical perspective, similar forms have been used in scalar field dynamics to phenomenologically reproduce late-time acceleration (e.g., see hyperbolic scalar potentials or effective field models in quintessence scenarios). Although our model does not invoke a specific Lagrangian potential, the use of a tanh-profile energy density provides a controlled and physically interpretable evolution without committing to a unique scalar field theory. Lastly, by working directly with the energy density as a redshift function, we bypass uncertainties related to the field’s dynamics and instead focus on its observable imprint on cosmic expansion. This approach is particularly advantageous in observational cosmology, where quantities like $\rho(z)$ are often more directly constrained than the underlying microphysics. Related strategies—where energy densities are parameterized rather than derived—have been adopted in various works, underscoring the practicality of such methods in dark energy reconstruction \cite{GEf99,Feng11,Ian20,Wang23}.

Starting from the modified Friedmann equation (\ref{20}) and incorporating the redshift-dependent expression for the scalar field energy density along with the standard matter density evolution, we obtain an expression for the Hubble parameter as a function of redshift. This takes the form:
\begin{equation}\label{33}
H^{2}=-\frac{\delta}{6\beta}\left[\rho_{m0}(1+z)^{3}+\rho_{c0}tanh(A+Bz)\right],
\end{equation} 
To characterize the expansion dynamics more conveniently, we define a dimensionless Hubble parameter, normalized with respect to the present Hubble rate. This approach introduces the density contrast with respect to the critical value necessary for spatial flatness. The critical energy density, given by $\rho_{c}=3H^{2}$, where the density parameter for this is defined as: $\Omega=\frac{\rho}{\rho_{c}}$, sets the benchmark for comparing contributions from various cosmic components.

Substituting the present-day values into the expression for the Hubble parameter, we evaluate equation (\ref{33}) at redshift $z=0$, yielding an expression normalized by the present Hubble constant $H_{0}$. To facilitate comparison with observations, we introduce the matter density parameter $\Omega_{m0}=\frac{\rho_{m0}}{3H_{0}^{2}}$, which quantifies the current matter contribution relative to the critical density. This leads to the dimensionless form:
\begin{equation}\label{34}
H^{2}=-\frac{\delta H_{0}^{2}}{2\beta}\left[\Omega_{m0}(1+z)^{3}+tanh(A+Bz)\right],
\end{equation}
To proceed further, we extract the ratio $-\frac{\delta}{2\beta}$ from equation (\ref{34}), which governs the overall expansion dynamics. Using the normalized Hubble parameter relation at present time, we impose the condition that $\frac{H^{2}}{H_{0}^{2}}=1$. Substituting into equation (\ref{34}), we obtain:
\begin{equation}\label{35}
\frac{\delta}{2\beta}=-\frac{1}{\Omega_{m0}+tanh(A)},
\end{equation}
To fix this ratio and reduce the number of independent parameters in the model, we now impose a consistency condition. Since the total normalized energy density today must satisfy $\Omega_{m0}+\Omega_{\phi 0}=1$. It leads to the explicit expression: $A=tanh^{-1}(1-\Omega_{m0})$. Substituting this back into the expression (\ref{35}), we derive $\frac{\delta}{2\beta}=-1$. Thus, the ratio $\frac{\delta}{2\beta}$ becomes uniquely determined, providing a valuable constraint that ensures the model aligns with the present-day expansion rate. This condition not only enforces physical consistency at $z=0$ but also simplifies the parametric structure of the model by removing one degree of freedom.

To ensure consistency of the model with present-time observations, we proceed by selecting representative numerical values for the parameters $\beta$ and $\delta$ that satisfy this relation. In particular, we choose: $\beta=0.6$ and $\delta=-1.2$, which clearly obey the condition $\frac{\delta}{2\beta}=-1$. This choice is motivated by the need to assign concrete values to the parameters in order to carry out numerical analysis and generate plots for key cosmological quantities such as the deceleration parameter, energy density, pressure and the scalar field's equation of state. These values are not arbitrary but are chosen to lie within a reasonable range that produces physically acceptable evolution behavior, in agreement with current cosmological constraints.

Thus, we arrive at a more compact and physically meaningful form for $H(z)$ as:
\begin{equation}\label{36}
H(z)=H_{0}\sqrt{\Omega_{m0}(1+z)^{3}+tanh[tanh^{-1}(1-\Omega_{m0})+Bz]}
\end{equation}
This form encapsulates the combined contributions of cold matter and the hyperbolically parameterized scalar field to the cosmic expansion. Importantly, it contains only three independent parameters $(H_{0}, \Omega_{m0}, B)$, which allows for a tractable confrontation with observations.
\section{Constraining the model with observational data and statistical analysis}\label{sec5}
\hspace{0.5cm} To constrain the key model parameters—namely, the present Hubble constant $H_{0}$, the present-day matter density parameter $\Omega_{m0}$ and the hyperbolic deformation parameter $B$, we perform a Bayesian statistical analysis using current observational datasets. Specifically, we utilize: Cosmic Chronometer (CC) Hubble parameter measurements $H(z)$, Baryon Acoustic Oscillations (BAO) and the Pantheon+ compilation of Type Ia supernovae. We implement the Markov Chain Monte Carlo (MCMC) sampling technique using the emcee Python package to explore the posterior distributions of the parameters and identify their best-fit values. The likelihood function is assumed to follow the standard Gaussian form: $\mathcal{L}\propto e^{-\frac{\chi^{2}_{total}}{2}}$, where the total chi-square, $\chi^{2}_{total}$ is constructed by summing over the contributions from each dataset. This approach allows us to simultaneously fit the model against multiple probes of the cosmic expansion history. We adopt the following flat priors for the parameters: $H_{0}=[60,100]$, $\Omega_{m0}=[0,1]$ and $B=[-1,1]$.
\subsection{Constraints from CC data}\label{sec5.1}
\hspace{0.5cm} To gauge the effectiveness of our $f(Q,L_{m})$ model, we employ cosmic chronometer data, yielding model-independent $H(z)$ values from the differential ages of galaxies with minimal star formation. We analyze $31$ observational data points over a range of redshifts, previously documented in \cite{Moresco16,AS25,NM23}. These measurements offer significant value in reconstructing the cosmic expansion history and are a vital component of our model's statistical framework. We construct the corresponding chi-square function in the following form:
\begin{equation}\label{37}
\chi^{2}_{CC}=\sum_{l=1}^{31}\frac{\big[H(z_{l},H_{0},\Omega_{m0},B)-H_{obs}(z_{l})\big]^{2}}{\sigma^{2}(z_{l})},
\end{equation}
The Hubble parameter's theoretical value $H(z_{l},H_{0},\Omega_{m0},B)$ at redshift $z_{l}$ is determined by the parameters $(H_{0}, \Omega_{m0}, B)$, $H_{obs}(z_{l})$ is the observed Hubble parameter and $\sigma^{2}(z_{l})$ denotes the corresponding observational uncertainty.
\subsection{BAO data}\label{sec5.2}
\hspace{0.5cm} BAO reflect the residual patterns of sound waves in the early Universe's plasma. With the Universe's expansion and cooling, the oscillations became embedded in the large-scale structure, yielding a fixed comoving scale called the sound horizon. This scale, approximately $r_{d}\approx147 Mpc$ functions as a standard ruler for cosmological distance measurements. Our analysis includes a thorough collection of $15$ BAO data points and recent DESI DR2 BAO dataset, sourced from recent galaxy redshift surveys and Lyman-$\alpha$ forest observations. These observations constrain cosmic expansion by benchmarking theoretical distance measures against the BAO scale, utilizing three crucial ratios \cite{DESI24,AG25,AGA25,Karim2025}. This ratio quantifies the relationship between angular separations on the sky and is defined as the transverse comoving distance ratio, which equals:
\begin{equation}\label{38}
\frac{d_{M}(z)}{r_{d}}=\frac{D_{L}(z)}{r_{d}(1+z)},
\end{equation}
The Hubble distance ratio quantifies the expansion rate along our line of sight and is given by the formula:
\begin{equation}\label{39}
\frac{d_H(z)}{r_d} = \frac{c}{r_d H(z)}.
\end{equation}
The volume-averaged distance ratio provides an effective isotropic measurement by averaging over both radial and transverse directions, expressed as
\begin{equation}\label{40}
\frac{d_V(z)}{r_d} = \frac{\left[z d_M^2(z) d_H(z)\right]^{1/3}}{r_d}.
\end{equation}
To test the model's fit to BAO data, we build the chi-square function in the following form:
\begin{equation}\label{41}
\chi^{2}_{BAO}=\sum_{k=1}^{N} \bigg[\frac{Y_{k}^{th}-Y_{k}^{obs}}{\sigma_{Y_{k}}}\bigg]^{2},
\end{equation}
where $Y_{k}^{th}$ and $Y_{k}^{obs}$ denote the theoretical and measured values of the BAO observable at redshift $z_{k}$ and $\sigma_{Y_{k}}$ represents the corresponding uncertainties. This approach allows us to constrain the cosmological parameters by quantifying the degree to which the model reproduces the observed structure of the Universe.
\subsection{Type Ia Supernovae and the Pantheon+ dataset}\label{sec5.3}
\hspace{0.5cm} Type Ia supernovae are exceptional cosmic distance indicators, providing a consistent benchmark for measuring the Universe's expansion. The observed quantity of interest is the distance modulus, defined as the logarithmic difference between the observed flux and the intrinsic luminosity. The distance modulus in a spatially flat universe is calculated from the luminosity distance $d_{L}(z)$ which is derived from integrating the inverse of the Hubble parameter with respect to redshift:
\begin{equation}\label{42}
\mu(z)=5\log_{10} \left[\frac{d_L(z)}{\text{Mpc}}\right] + 25,
\end{equation}
where
\begin{equation}\label{43}
d_L(z)=c(1+z)\int_0^z \frac{dz'}{H(z'; H_0, \Omega_{m0}, B)}.
\end{equation}
Our analysis utilizes the Pantheon$+$ dataset, featuring $1701$ SNe Ia measurements across the redshift range $0.001<z<2.3$. The dataset delivers tight constraints on cosmological parameters, especially for low and intermediate redshift ranges. To determine the agreement between theory and observation, we compute the chi-square statistic, defined as:
\begin{equation}\label{44}
\chi^2_{SNe}(H_0, \Omega_{m0}, B)=\sum_{i=1}^{1701}\frac{[\mu_{th}(z_i; H_0, \Omega_{m0}, B)-\mu_{obs}(z_i)]^2}{\sigma^2(z_i)},
\end{equation}
with $\mu_{th}$ is the theoretical distance modulus, $\mu_{obs}$ represents the observed value and $\sigma^2(z_i)$ denotes the uncertainty at redshift $z_{i}$.
\subsection{Combined statistical analysis and parameter estimation}\label{sec5.4}
\begin{figure}[hbt!]
  \centering
  \includegraphics[scale=0.33]{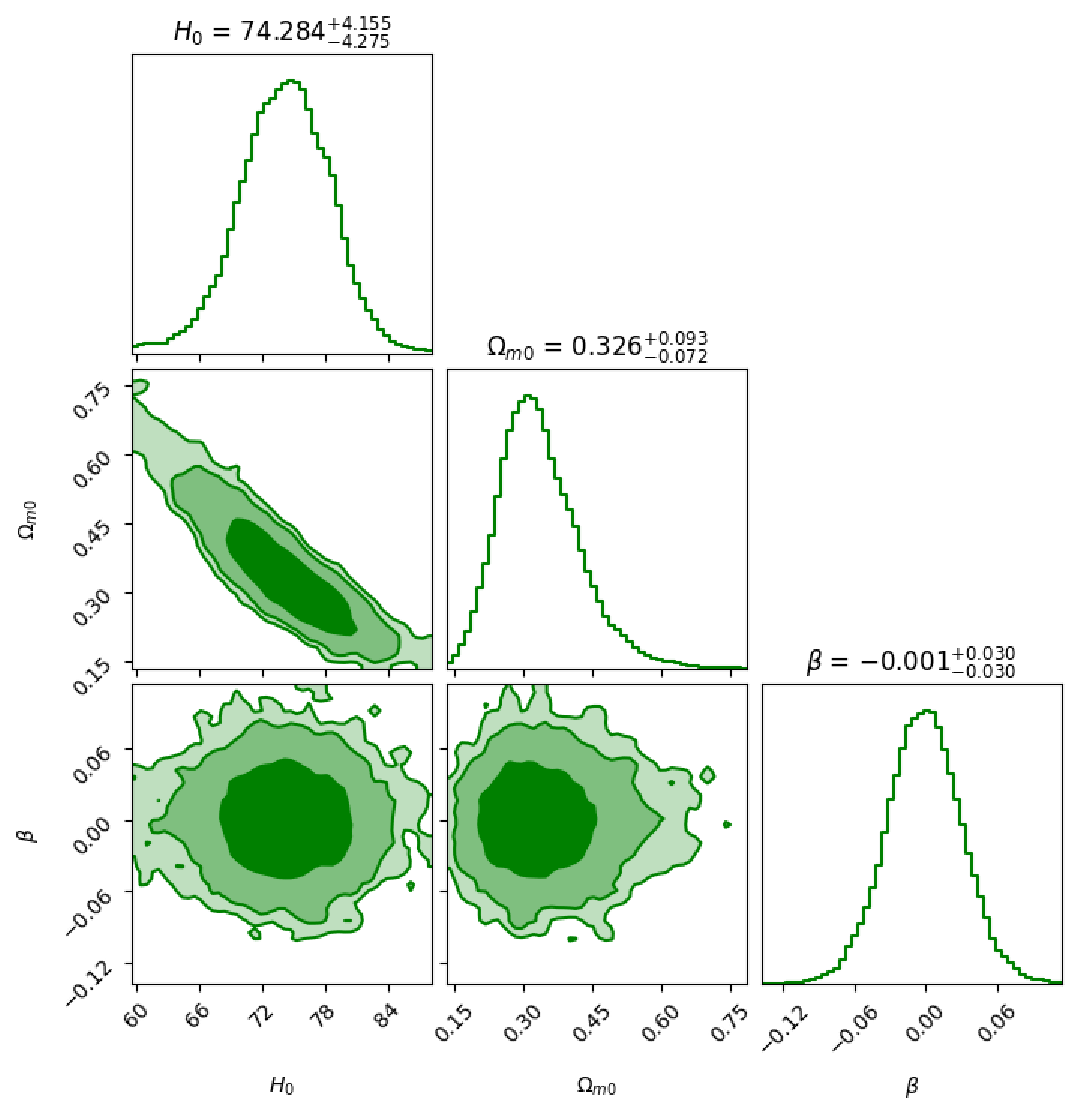}
  \caption{Joint estimation of $(H_0, \Omega_{m0}, B)$ with confidence level visualization. The shaded regions denote the $1\sigma$ $(68.27\%)$, $2\sigma$ $(95.45\%)$ and $3\sigma$ $(99.73\%)$ confidence levels.}\label{fig:f1}
\end{figure}

\begin{figure}[hbt!]
  \centering
  \includegraphics[scale=0.33]{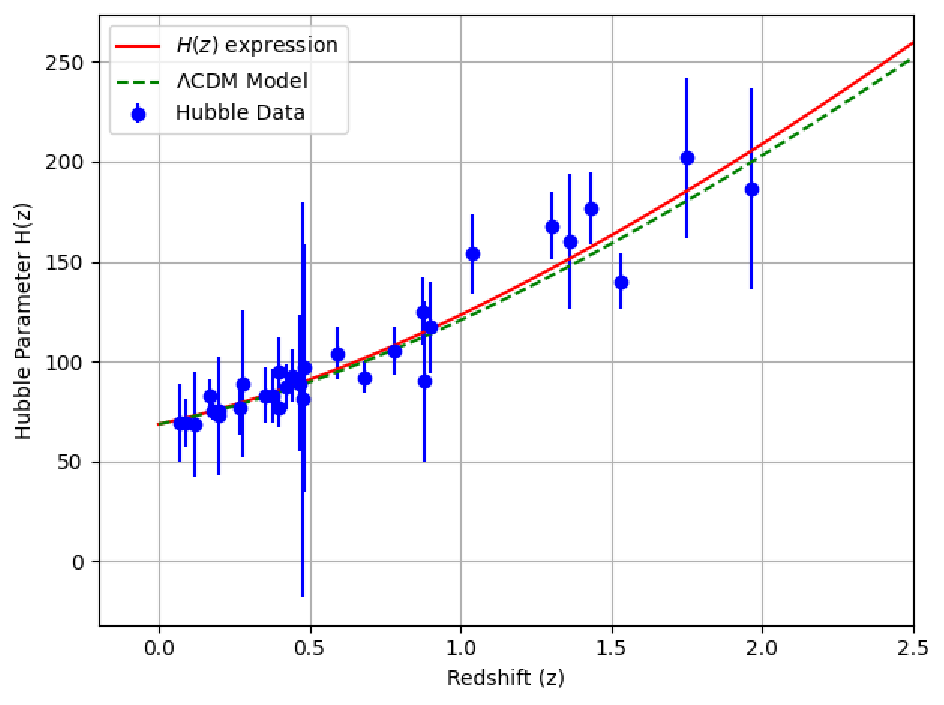}~
  \includegraphics[scale=0.33]{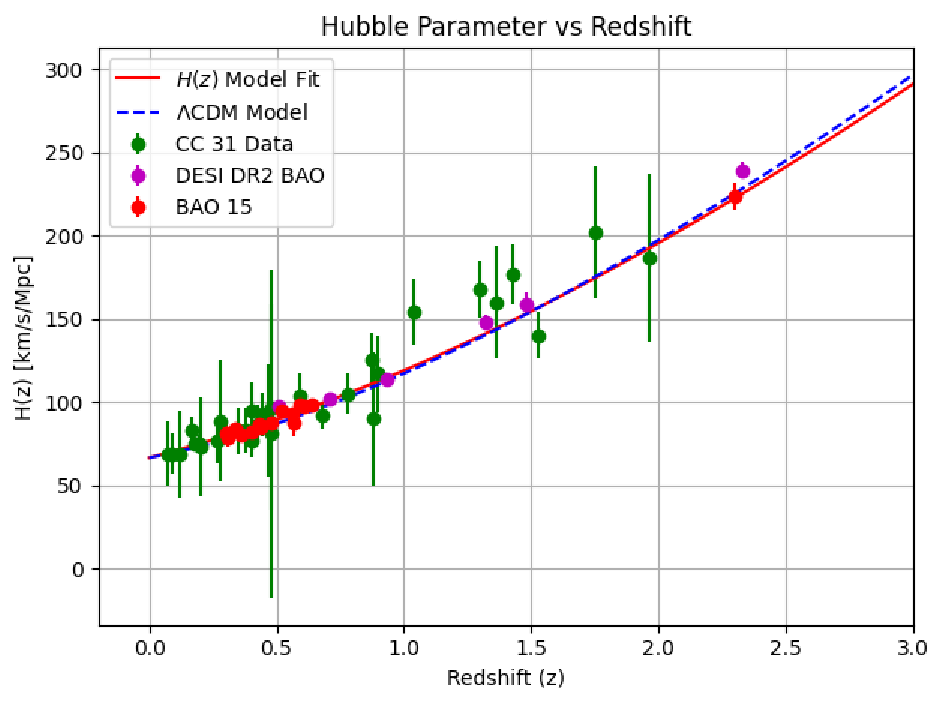}~~
  \includegraphics[scale=0.28]{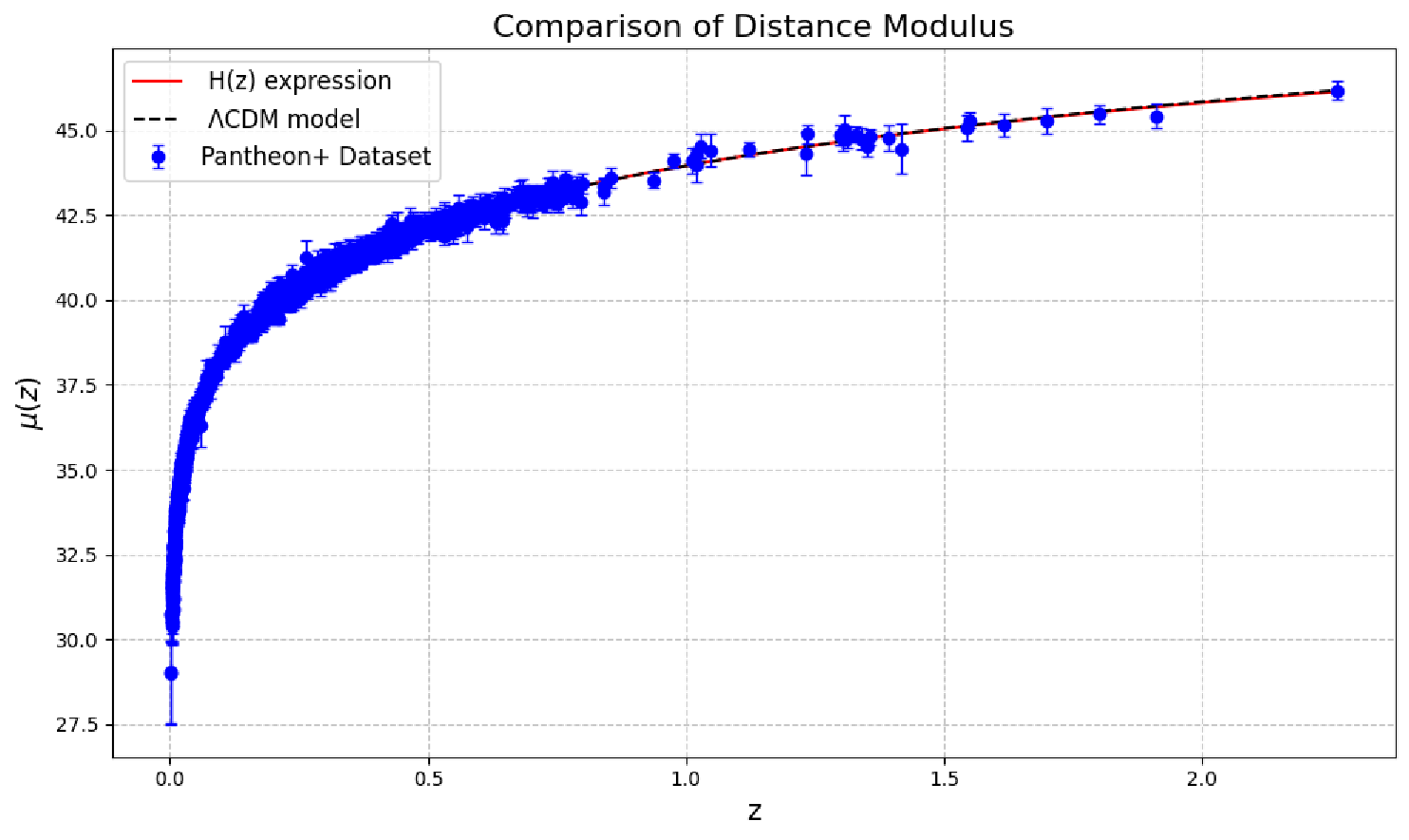}
  \caption{Error bar comparison illustrating the variability in parameter estimates across different dataset combinations.}\label{fig:f2}
\end{figure}

\hspace{0.5cm} Our joint analysis of multiple datasets, including CC, BAO (including DESI DR2) and Pantheon+, yields robust insights into the cosmological model. The combination of datasets offers complementary views of the Universe's expansion history, leading to a significant improvement in parameter estimation precision. The joint chi-square function for the combined analysis is expressed as:
\begin{equation}\label{45}
\chi^{2}_{total}=\chi^{2}_{CC}+\chi^{2}_{BAO}+\chi^{2}_{SNe Ia},
\end{equation}
The MCMC method is used to probe the parameter space $(H_0, \Omega_{m0}, B)$. This probabilistic method samples from the posterior distributions of the parameters by maximizing the combined likelihood derived from the total chi-square function. The resulting constraints are visualized through two-dimensional confidence contours and one-dimensional marginalized distributions. These plots provide insights into both the most probable values of the parameters and their correlations. The combined dataset analysis significantly reduces degeneracies and narrows down the allowed parameter space, offering tighter bounds compared to individual datasets alone. Figures \ref{fig:f1} and \ref{fig:f2} display the contour plot and posterior distribution, illustrating the compatibility of the different observational probes and reinforcing the effectiveness of a multi-dataset approach in cosmological parameter estimation.

The estimated parameters of the $f(Q,L_{m})$ gravity model are determined to be: $H_{0}=74.284^{+4.155}_{-4.275}$, $\Omega_{m0}=0.326^{+0.093}_{-0.072}$ and $B=-0.001^{+0.030}_{-0.030}$. The Hubble constant $H_{0}$ we obtain is consistent with local measurements like SH0ES $H_{0}\sim 73.2\pm1.3$ km/s/Mpc \cite{Adam21} and overlaps with Planck CMB estimates $H_{0}\sim 67.4\pm0.5$ km/s/Mpc \cite{Planck20} within $1.5\sigma-2\sigma$ bounds. This suggests that the $f(Q,L_{m})$ model could provide a potential resolution to the Hubble tension by accommodating a higher local expansion rate. The present matter density parameter $\Omega_{m0}\approx0.327$ is consistent with recent large-scale structure and Planck measurements, indicating that the matter content inferred by the model is in agreement with the standard cosmological observations. The scalar field parameter $B\approx-0.001$ with symmetric error bars indicates a value statistically consistent with zero within $1\sigma$ uncertainty. This suggests that the contribution of the scalar field correction to the gravitational action is minimal in the current epoch. However, the non-zero central value still allows for non-trivial effects at early or late times, which could influence cosmic acceleration or structure formation subtly \cite{Aub15,Chen11,Scap20}.

The $f(Q,L_{m})$ gravity model, based on the derived parameter values, shows strong consistency with current cosmological datasets. It effectively describes the late-time acceleration, fits both local and high-redshift data and offers a viable solution to the Hubble tension. This confirms the model's success in capturing key features of the Universe’s evolution.
\section{Cosmic evolution analysis}\label{sec6}
\hspace{0.5cm} This section explores the dynamics of important cosmological parameters in the $f(Q,L_{m})$ gravity. We analyze the evolution of the deceleration parameter $q$, energy density $\rho$, pressure $p$ and EoS parameter $\omega$, which together characterize the expansion history of the Universe. Additionally, we explore the dynamics of the scalar field $\phi^{2}$ and its potential $V(\phi)$, which offers insights into the role of the scalar sector in driving the observed cosmic acceleration. These quantities collectively provide a comprehensive understanding of the model's compatibility with observational cosmology.
\subsection{Deceleration parameter and cosmic acceleration transition}\label{sec6.1}
\hspace{0.5cm} Understanding the deceleration parameter $q$ is essential for unraveling the mysteries of cosmic expansion and its acceleration. The expression of it is
\begin{equation}\label{46}
q=-1-\frac{\dot{H}}{H^{2}},
\end{equation}
The deceleration parameter $q$ is positive for decelerating expansion and negative for accelerating expansion. Observational evidence from Type Ia supernovae, CMB and BAO data confirms that the Universe has transitioned from a decelerated phase in the past to an accelerated expansion at low redshifts. The transition redshift $z_{tr}$ typically falls in the range $0.5<z_{tr}<1$ and the present-day value of the deceleration parameter is constrained around $q_{0}\approx-0.5$ to $-0.6$, consistent with a $\Lambda$CDM-like evolution. In our model, we can estimate the deceleration parameter as an expression of redshift $z$ via equation (\ref{36}).
\begin{equation}\label{47}
q=-1+\frac{(1+z)\left[3\Omega_{m0}(1+z)^{2}+Bsech^{2}\{tanh^{-1}(1-\Omega_{m0})+Bz\}\right]}{2\left[\Omega_{m0}(1+z)^{3}+tanh\{tanh^{-1}(1-\Omega_{m0})+Bz\}\right]},
\end{equation}
\begin{figure}[hbt!]
  \centering
  \includegraphics[scale=0.4]{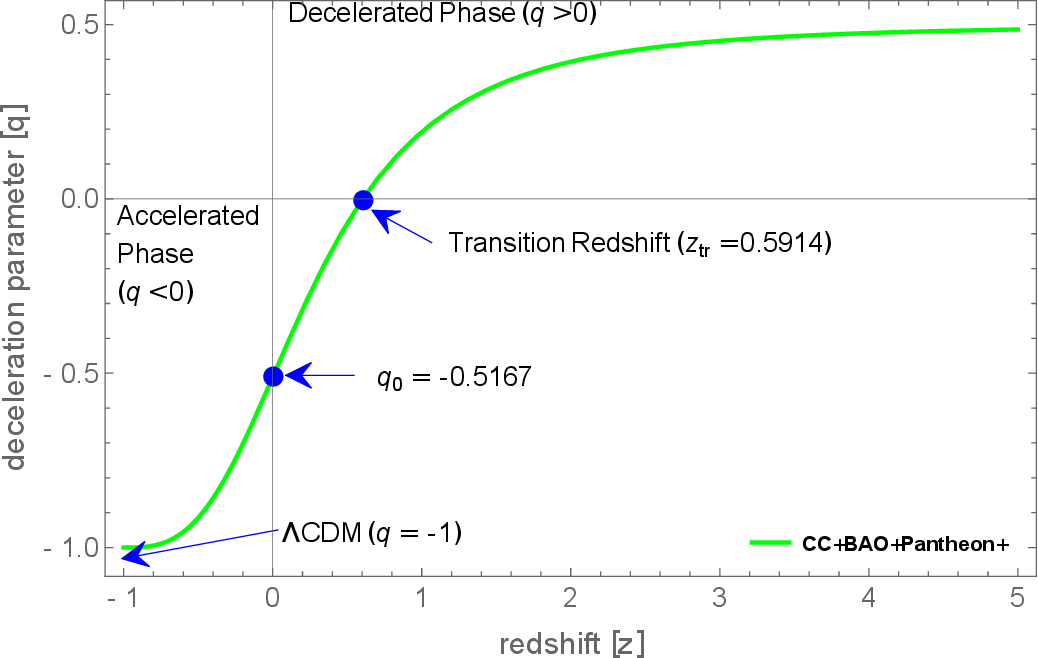}
  \caption{$q(z)$ vs. redshift for the best-fit parameters.}\label{fig:f3}
\end{figure}

The $q(z)$ evolution is depicted in Figure \ref{fig:f3}, utilizing the optimal model parameters. The plot shows that at high redshift (early Universe), the deceleration parameter begins near $q\approx0.5$, consistent with a matter-dominated decelerated phase. The Universe undergoes a transition from deceleration to acceleration at a redshift of $z_{tr}=0.5914$, which lies well within the observational range. As $z\rightarrow-1$ (future), the curve asymptotically approaches $q=-1$ indicating that the model mimics a de Sitter expansion akin to the $\Lambda$CDM scenario. The present-day value is found to be $q_{0}=-0.5167$, which agrees remarkably well with current observational estimates. The results show that our model accurately captures the expected behavior of cosmic acceleration. The derived transition redshift and current deceleration parameter are consistent with recent observational bounds, reinforcing the validity of the $f(Q,L_{m})$ gravity framework in describing the late-time dynamics of the Universe. 
\subsection{Energy density and pressure evolution}\label{sec6.2}
\begin{figure}[hbt!]
  \centering
  \includegraphics[scale=0.3]{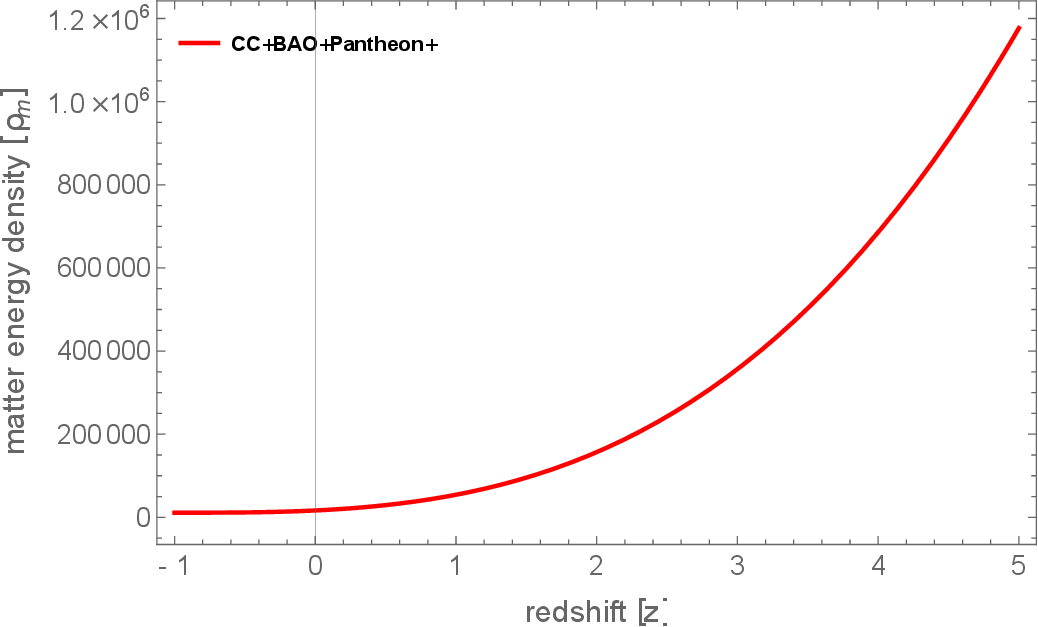}~~
  \includegraphics[scale=0.3]{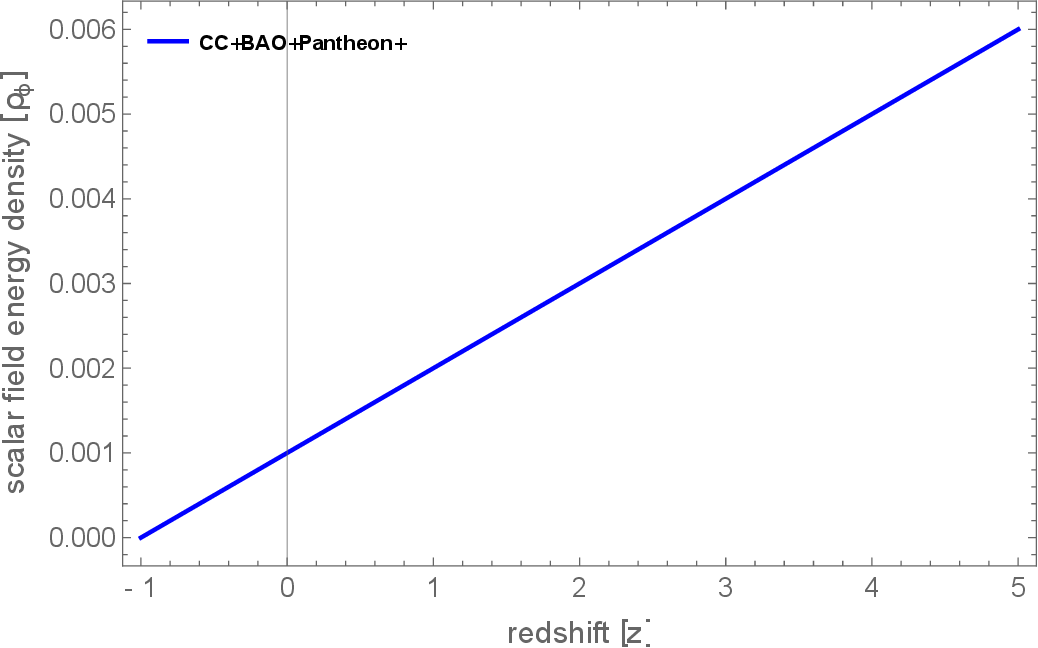}~~
  \includegraphics[scale=0.3]{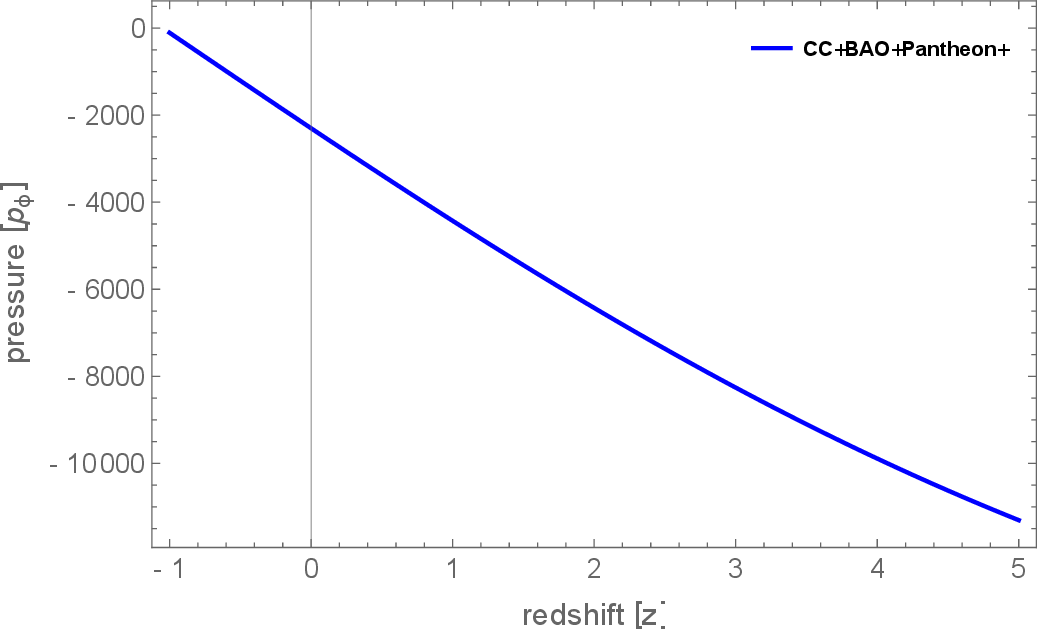}
  \caption{Energy densities and scalar field pressure as functions of redshift for the parameter $B=1$, $\beta=0.6$ and $\delta=-1.2$.}\label{fig:f4}
\end{figure}

\hspace{0.5cm} The dynamical behavior of the Universe is significantly influenced by the evolution of energy density and pressure. In modified gravity frameworks like $f(Q,L_{m})$ both matter and scalar field contributions must be analyzed separately to trace their influence on cosmic expansion. While matter follows the standard continuity equation, the scalar field’s dynamics are governed by the modified field equations, resulting in distinct behavior in its energy density and pressure. Observational consistency demands that energy densities remain positive and that pressure-particularly from dark energy components-becomes negative to drive acceleration. Using equations (\ref{20}), (\ref{21}), (\ref{32}) and (\ref{36}), we calculate the matter energy density $\rho_{m}$, scalar field energy density $\rho_{\phi}$ and scalar field pressure $p_{\phi}$. These quantities are plotted in Figure \ref{fig:f4} to analyze their redshift dependence under the best-fit parameter values.
\begin{equation}\label{48}
\rho_{m}=-\frac{6\beta H_{0}^{2}}{\delta}\bigg(\Omega_{m0}(1+z)^{3}+tanh[tanh^{-1}(1-\Omega_{m0})+Bz]\bigg)-tanh[tanh^{-1}(1-\Omega_{m0})+Bz],
\end{equation}
\begin{equation}\label{49}
\rho_{\phi}=tanh\left(tanh^{-1}(1-\Omega_{m0})+Bz\right),
\end{equation}
and
\begin{equation}\label{50}
p_{\phi}=\frac{2\beta H_{0}^{2}}{\delta}\bigg(B(1+z)sech^{2}[tanh^{-1}(1-\Omega_{m0})+Bz]+3tanh[tanh^{-1}(1-\Omega_{m0})+Bz]\bigg).
\end{equation}

From Figure \ref{fig:f4}, we observe that both matter and scalar field energy densities remain strictly positive throughout the cosmic evolution, satisfying physical viability. The matter energy density $\rho_{m}(z)$ begins with a high positive value at early times (large $z$) and gradually decays, approaching zero in the far future, as expected from standard cosmology. In contrast, the scalar field energy density $\rho_{\phi}(z)$ starts from a small positive value and similarly decreases toward zero at late times. This behavior signifies that the scalar field contribution becomes dynamically negligible in the far future but plays a crucial role during intermediate redshifts where cosmic acceleration is active.

For the pressure component $p_{\phi}(z)$, the plot reveals a consistently negative trend across redshifts. This negative pressure is a key feature responsible for driving the late-time acceleration and mimics the behavior of dark energy or a cosmological constant. The pressure approaching more negative values aligns with the observed acceleration and the asymptotic de Sitter behavior in our model.
\subsection{Equation of state parameter and dark energy dynamics}\label{sec6.3}
\begin{figure}[hbt!]
  \centering
  \includegraphics[scale=0.4]{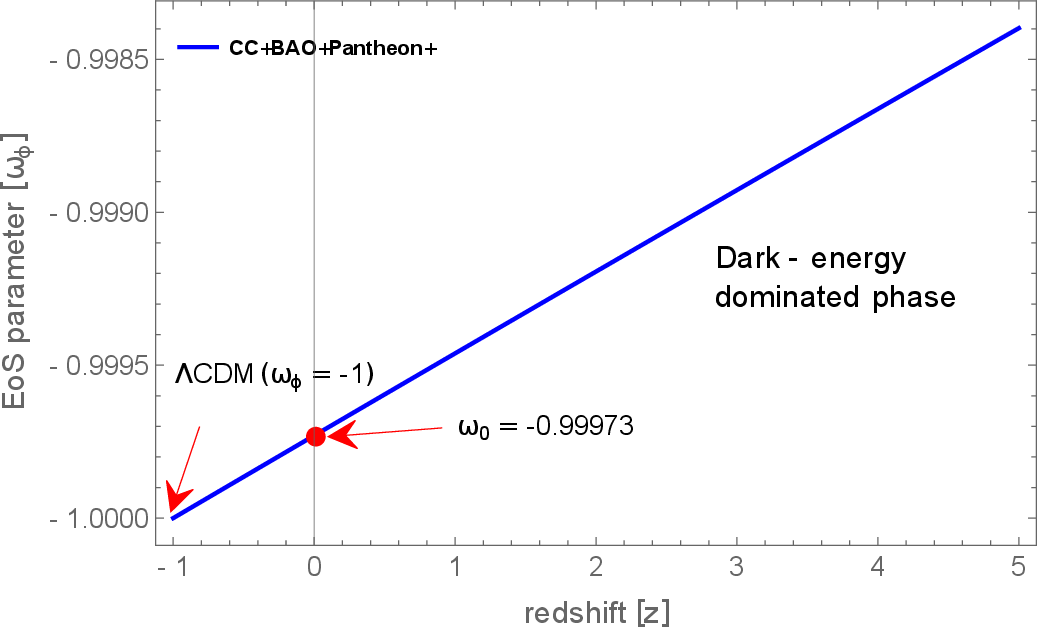}
  \caption{The changing profile of the scalar field EoS parameter $\omega_{\phi}(z)$ over redshift for $B=1$, $\beta=0.6$ and $\delta=-1.2$.}\label{fig:f5}
\end{figure}

\hspace{0.5cm} The EoS parameter $\omega=\frac{p}{\rho}$ expresses how the pressure of a cosmological fluid is connected to its energy density. In the case of the scalar field $\phi$, the EoS parameter $\omega_{\phi}$ provides critical insights into the nature of dark energy and its role in the Universe’s expansion. Observationally, dark energy is best described by a nearly constant EoS parameter close to $\omega\approx-1$, consistent with a cosmological constant ($\Lambda$CDM). Current data from Planck $2018$, Supernovae Type Ia and BAO suggest that the present-day value lies within $-1.03<\omega_{0}<-0.97$. Any deviation from $\omega=-1$ can indicate dynamical dark energy, often classified into two regimes: Quintessence phase $-1<\omega<-0.33$ and phantom phase $\omega<-1$. From our model, the scalar field EoS parameter $\omega_{\phi}$ is calculated using equations (\ref{49}) and (\ref{50}). The behavior of $\omega_{\phi}$ is plotted in Figure \ref{fig:f5} based on the best-fit values.
\begin{equation}\label{51}
\omega_{\phi}=-1+\frac{B(1+z)sech[tanh^{-1}(1-\Omega_{m0})+Bz]}{3tanh[tanh^{-1}(1-\Omega_{m0})+Bz]}.
\end{equation}

As illustrated in Figure \ref{fig:f5}, the EoS parameter $\omega_{\phi}(z)$ starts at a value of $\omega_{\phi}(z>>1)=-0.9948$, which indicates a near-$\Lambda$CDM behavior during earlier times. As redshift decreases, $\omega_{\phi}$ slowly transitions into a more negative regime, ultimately converging to $\omega_{\phi}(z\rightarrow-1)\rightarrow-1$, signifying a de Sitter phase in the far future. The present-day value is found to be $\omega_{0}=-0.99973$, which lies comfortably within observational limits and further confirms the viability of our scalar field description. This evolution shows that the scalar field behaves very closely to a dynamical dark energy model in the early and intermediate epochs, but asymptotically behaves like a cosmological constant. There is no crossing of the phantom divide $\omega<-1$, suggesting that the model avoids instabilities typically associated with phantom fields.
\subsection{Scalar field behavior and potential evolution}\label{sec6.4}
\hspace{0.5cm} In scalar field cosmology, the scalar field $\phi$ and its associated potential $V(\phi)$ are fundamental in modeling the accelerated expansion of the Universe. The dynamics of $\phi$ determine the energy density and pressure, while the potential $V(\phi)$ encapsulates the self-interacting nature of the field. A well-behaved scalar field model typically requires: $\dot{\phi}^{2}>0$ for physical viability and a positive and decreasing potential $V(\phi)$, especially in late-time cosmic acceleration, to ensure a smooth transition to a de Sitter phase. In $f(Q,L_{m})$ gravity, the scalar field emerges as an effective component capturing dark energy-like behavior. The evolution of $\dot{\phi}^{2}$ and $V(\phi)$ can thus be used to further validate the model’s consistency with realistic cosmic evolution. Our model leads to explicit forms for the $\dot{\phi}^{2}$ and $V(\phi)$ as given in equations (\ref{20}), (\ref{21}), (\ref{32}) and (\ref{36}).
\begin{eqnarray}\label{52}
\dot{\phi}^{2}&=&tanh[tanh^{-1}(1-\Omega_{m0})+Bz]+\frac{2\beta H_{0}^{2}}{\delta}\bigg(B(1+z)sech^{2}[tanh^{-1}(1-\Omega_{m0})+Bz]\\\nonumber
&&+3tanh[tanh^{-1}(1-\Omega_{m0})+Bz]\bigg),
\end{eqnarray}
\begin{eqnarray}\label{53}
V(\phi)&=&tanh[tanh^{-1}(1-\Omega_{m0})+Bz]-\frac{2\beta H_{0}^{2}}{\delta}\bigg(B(1+z)sech^{2}[tanh^{-1}(1-\Omega_{m0})+Bz]\\\nonumber
&&+3tanh[tanh^{-1}(1-\Omega_{m0})+Bz]\bigg).
\end{eqnarray}
\begin{figure}[hbt!]
  \centering
  \includegraphics[scale=0.4]{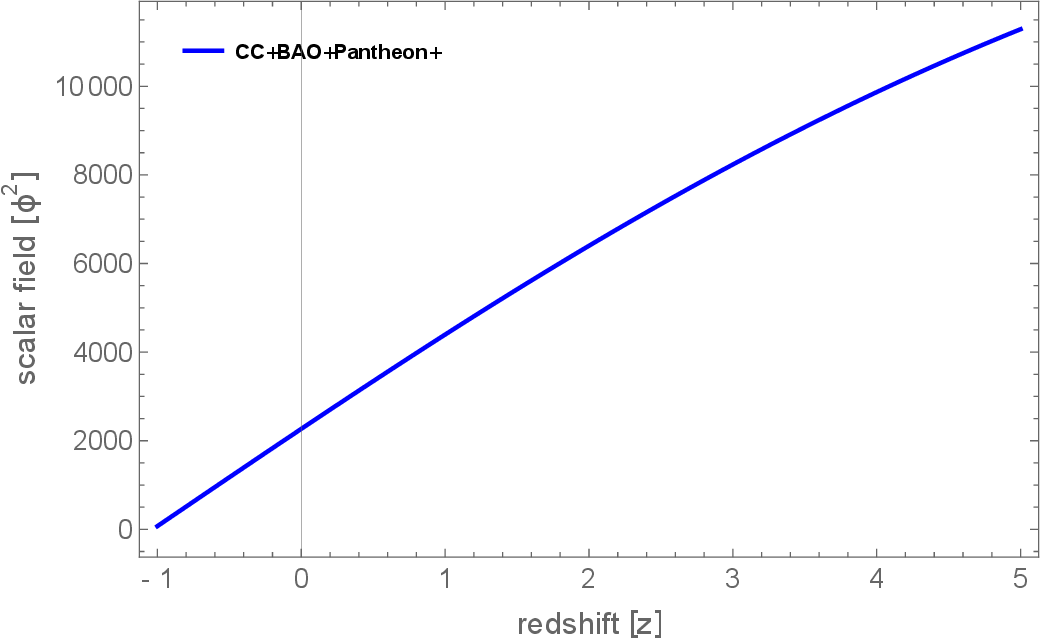}~~
  \includegraphics[scale=0.4]{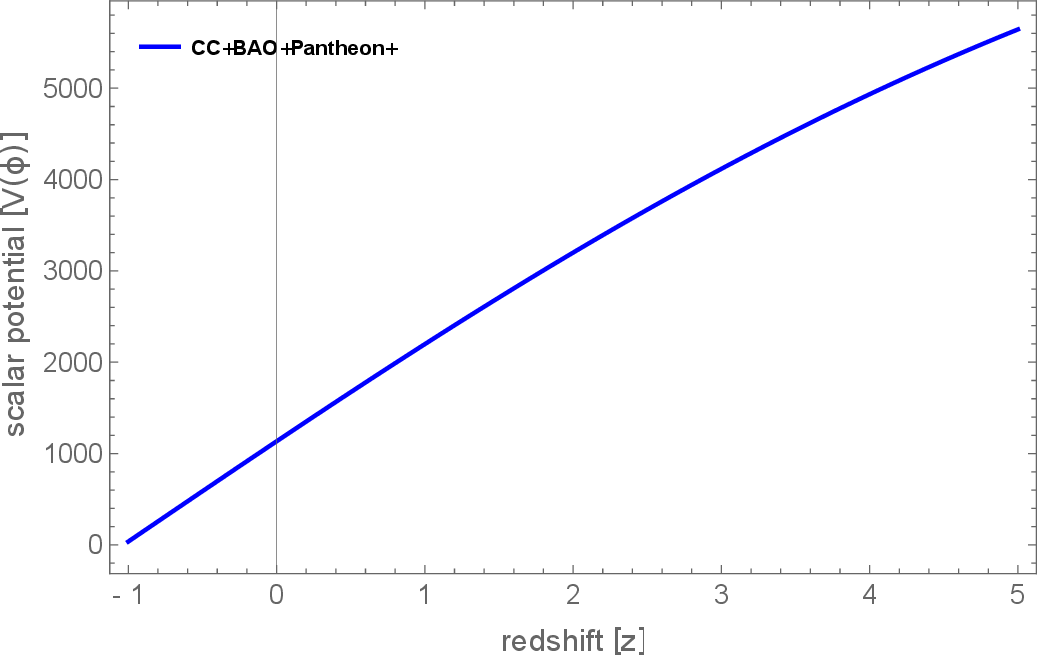}
  \caption{Variation of the $\dot{\phi}^{2}$ and $V(\phi)$ as functions of redshift.}\label{fig:f6}
\end{figure}

The progression of $\dot{\phi}^{2}$ and $V(\phi)$ over redshift is visualized in Figure \ref{fig:f6}. The plots clearly show that: $\dot{\phi}^{2}$ remains strictly positive across the entire redshift range, ensuring the physical consistency of the field and the scalar potential $V(\phi)$ also exhibits a positive profile and gradually decreases as the Universe evolves, indicating a graceful evolution toward a stable, dark-energy–dominated state. This behavior aligns with the expected features of quintessence or effective $\Lambda$ models, where the field slowly rolls down its potential and drives cosmic acceleration. The monotonic decrease of $V(\phi)$ also signals that the field becomes dynamically less influential at late times, stabilizing the expansion.
\subsection{Evolution of density parameters}\label{sec6.5}
\hspace{0.5cm} The density parameters $\Omega_{m}$ and $\Omega_{\phi}$ describe the fractional contributions of matter and scalar field energy densities to the total energy budget of the Universe. These parameters evolve with redshift and provide a powerful tool to understand the cosmological dynamics at different epochs. In a viable cosmological model: $\Omega_{m}$ dominates in the early Universe $(z>>1)$ and $\Omega_{\phi}$ dominates at late times, leading to the accelerated expansion. The sum $\Omega_{m}+\Omega_{\phi}\approx1$ reflects spatial flatness, as supported by CMB observations. Current estimates suggest that $\Omega_{m0}\approx0.3$ and $\Omega_{\phi0}\approx0.7$, which any successful model should reproduce. The density parameters are constructed based on the energy densities derived from equations (\ref{48}) and (\ref{49}).
\begin{equation}\label{54}
\Omega_{m}(z)=\frac{\Omega_{m0}(1+z)^{3}}{\Omega_{m0}(1+z)^{3}+tanh[tanh^{-1}(1-\Omega_{m0})+Bz]},
\end{equation}
and 
\begin{equation}\label{55}
\Omega_{\phi}(z)=\frac{tanh[tanh^{-1}(1-\Omega_{m0})+Bz]}{\Omega_{m0}(1+z)^{3}+tanh[tanh^{-1}(1-\Omega_{m0})+Bz]}.
\end{equation}
\begin{figure}[hbt!]
  \centering
  \includegraphics[scale=0.4]{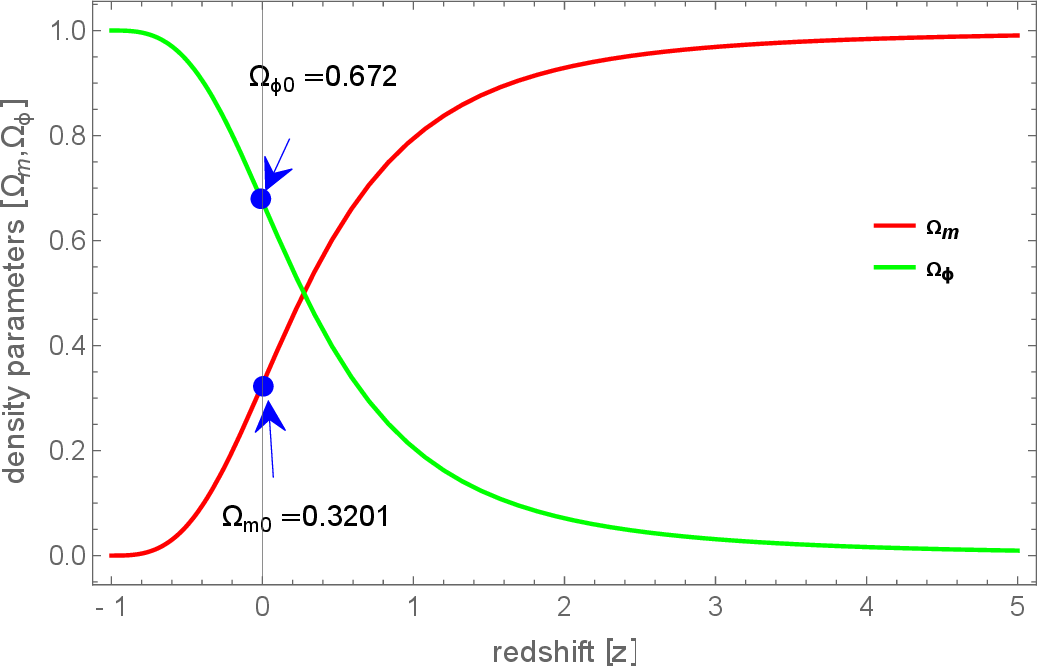}
  \caption{Tracking the redshift evolution of $\Omega_{m}$ and $\Omega_{\phi}$ for $B=1$.}\label{fig:f7}
\end{figure} 

We depict in Figure \ref{fig:f7} the changes in $\Omega_{m}(z)$ and $\Omega_{\phi}(z)$ through redshift. At high redshift $z>>1$, the $\Omega_{m}$ starts at unity, indicating a Universe dominated by matter in its early stages. As redshift decreases, $\Omega_{m}$ gradually decreases and tends to zero in the far future $(z\rightarrow-1)$. Conversely, the scalar field density parameter $\Omega_{\phi}$ begins near zero and increases over time, ultimately approaching unity. This transition illustrates the replacement of matter dominance with scalar field dominance which is consistent with the onset of late-time acceleration. The present-day values are found to be: $\Omega_{m0}=0.3201$ and $\Omega_{\phi0}=0.672$, which align closely with current observational data, reaffirming the model’s robustness and ability to describe the full evolution of the Universe.
\section{Cosmic age estimation}\label{sec7}
\hspace{0.5cm} The age of the Universe stands as a cornerstone parameter in cosmology, providing a crucial test for the consistency of any proposed model. It reflects the accumulated history of cosmic expansion and imposes a temporal limit on the formation of large-scale structures. High-precision measurements, especially from the cosmic microwave background (CMB) data—such as those reported by the Planck 2018 mission—estimate the Universe's current age to be around $13.8$ billion years within the context of the $\Lambda$CDM model. In modified gravity scenarios like $f(Q,L_{m})$, evaluating the cosmic age provides an independent check on the model’s consistency with observation. It is computed via the integral:
\begin{equation}\label{56}
t_{0}-t=\int_{0}^{z}\frac{dz}{(1+z)H(z)},
\end{equation}
which, when evaluated at $z=0$, gives the present age $t_{0}$. A viable model should yield a value close to this established benchmark. Substituting our expression for $H(z)$ from equation (\ref{36}) we compute the age numerically and compare it with observations.
\begin{equation}\label{57}
H_{0}(t_{0}-t)=\int_{0}^{z}\frac{dz}{(1+z)\sqrt{\Omega_{m0}(1+z)^{3}+tanh[tanh^{-1}(1-\Omega_{m0})+Bz]}},
\end{equation}
\begin{figure}[hbt!]
  \centering
  \includegraphics[scale=0.4]{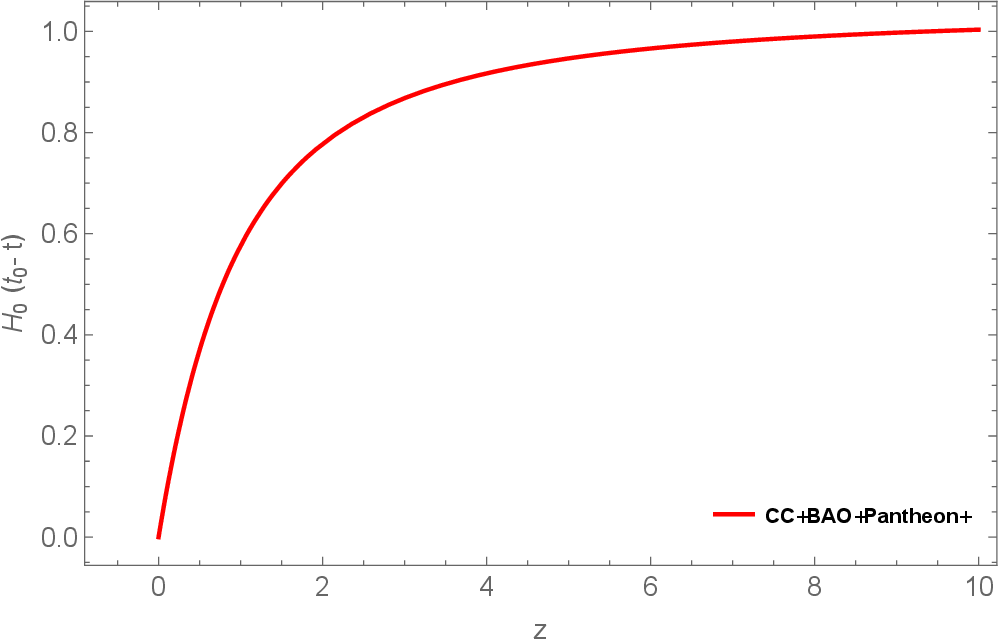}
  \caption{Computation of the age parameter $H_{0}t_{0}$ using the best-fit model parameters.}\label{fig:f8}
\end{figure} 

As shown in Figure \ref{fig:f8}, the computed value $H_{0}t_{0}=1.00382$ corresponds to a present cosmic age of about $t_{0}=13.51$ billion years. This result is remarkably consistent with the current astrophysical constraint from Planck, suggesting that our $f(Q,L_{m})$ model accurately reproduces the temporal evolution of the Universe. The closeness of this value to the $\Lambda$CDM prediction reinforces the observational viability of our model, especially under joint dataset constraints \cite{Cruz20,IB20}.
\section{Statefinder diagnostic and geometrical evolution}\label{sec8}
\hspace{0.5cm} The Statefinder diagnostic provides a valuable geometric approach for differentiating various dark energy models that extend beyond the conventional $\Lambda$CDM scenario. This method relies on two key parameters, $\{r, s\}$, derived from higher-order derivatives of the scale factor, making them effective indicators of the Universe's dynamical behavior \cite{Sahni03}.
\begin{equation}\label{58}
  r=\frac{\dddot a}{aH^{3}}=2q^{2}+q+(1+z)\frac{dq}{dz},
\end{equation}
\begin{equation}\label{59}
  s=\frac{(r-1)}{3(q-\frac{1}{2})}. \bigg(q\neq\frac{1}{2}\bigg).
\end{equation} 
These parameters act as a distinctive signature for each cosmological model, enabling precise identification based on their dynamic behavior. Different cosmological models trace unique trajectories in the $r$–$s$ plane. Chaplygin gas models often exhibit behavior with $r > 1$ and $s < 0$, indicating a strong deviation from $\Lambda$CDM at early times. In contrast, phantom and quintessence models typically lie in the region where $r < 1$ and $s > 0$, although the exact trajectory may vary depending on the specific dynamics \cite{Feng08,Wu07}. On the other hand, braneworld and various modified gravity theories, including our $f(Q, L_m)$ model, can display more complex or looping patterns in the $r$–$s$ plane, reflecting their richer geometric structure and evolution.
\begin{figure}[hbt!]
  \centering
  \includegraphics[scale=0.4]{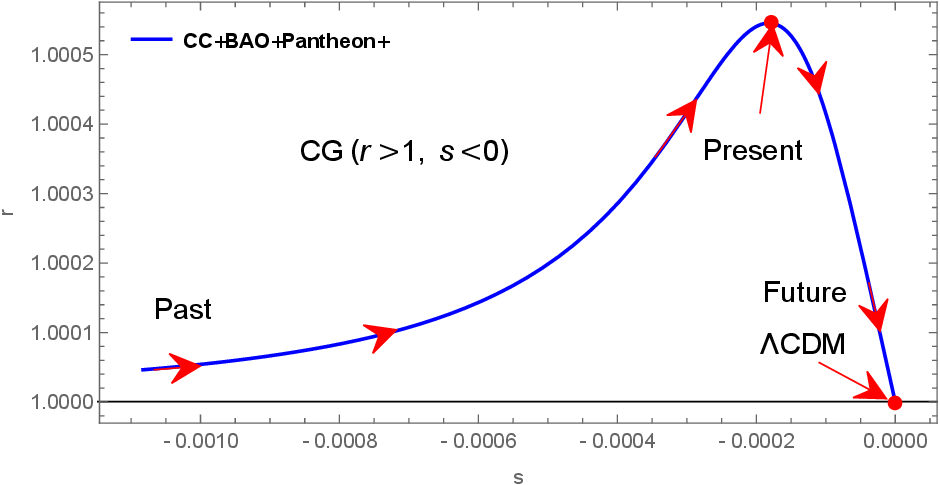}
  \caption{Evolution of the statefinder parameters.}\label{fig:f9}
\end{figure} 

From Figure \ref{fig:f9}, the trajectory in the $r$–$s$ plane reveals essential dynamical characteristics of our $f(Q, L_m)$ model. Initially, the model lies in the Chaplygin gas–like region, characterized by $r > 1$ and $s < 0$, reflecting a phase where the model’s effective dynamics differ notably from standard dark energy models. As the Universe evolves, the curve smoothly approaches a turning point near the present epoch at $r_0 = 1.00055$ and $s_0 = -0.00018$ which indicates the close agreement with current observational behavior. Eventually, the trajectory converges to the $\Lambda$CDM fixed point, $\{r, s\} = \{1, 0\}$, suggesting that our model asymptotically behaves like standard cosmology in the future. This continuous transition from a Chaplygin-like phase to a $\Lambda$CDM-like state reinforces the consistency and observational viability of our modified gravity model across different cosmic epochs \cite{Amits24}.
\section{Black hole accretion and mass evolution in $f(Q,L_{m})$ gravity}\label{sec9}
\hspace{0.5cm} Black hole accretion offers a powerful probe of the large-scale cosmic environment, particularly when dark energy is modeled by a dynamical scalar field. In this scenario, the black hole does not merely exist as an isolated astrophysical object but interacts with the evolving cosmic background. The accretion of matter and dark energy onto black holes can influence their mass evolution over time, making it a sensitive diagnostic of both gravitational theory and cosmological fluids. Unlike standard treatments that often assume negligible dark energy interaction with compact objects, in our model based on $f(Q,L_{m})$ gravity, we consider scalar field dark energy as a viable component of the cosmic fluid. The accretion process is governed by a generalized form of the Babichev equation, where the rate of black hole mass change is given by \cite{Babichev14}:
\begin{equation}\label{60}
\dot{M}=4\pi M^{2}A(\rho+p),
\end{equation}
with $\rho=\rho_{\phi}+\rho_{m}$ and $p=p_{\phi}$, as derived from equations (\ref{48}), (\ref{49}) and (\ref{50}). This formalism reflects the influence of the total effective fluid (both matter and scalar field components) on the black hole evolution.
\begin{equation}\label{61}
M=\frac{\delta}{8\pi AH_{0}\beta}\frac{1}{\displaystyle\int\bigg[\frac{Bsech^{2}[tanh^{-1}(1-\Omega_{m0})+Bz]-3\Omega_{m0}(1+z)^{2}}{\sqrt{\Omega_{m0}(1+z)^{3}+tanh[tanh^{-1}(1-\Omega_{m0})+Bz]}}\bigg]+I_{1}}.
\end{equation}

\begin{figure}[hbt!]
  \centering
  \includegraphics[scale=0.4]{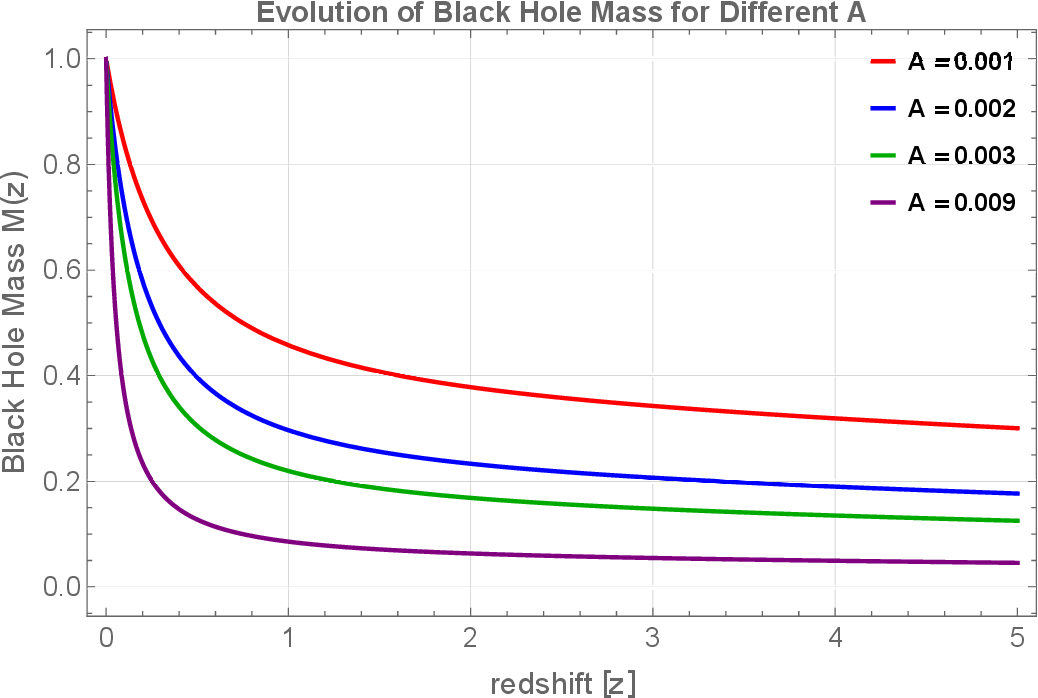}
  \caption{The variation of black hole mass $M(z)$ with redshift is explored for varying accretion constants $A$.}\label{fig:f10}
\end{figure}

To explore this scenario, we utilize the formulation of black hole mass dependent on redshift, as specified in equation (\ref{61}). In this equation, $I_{1}$ represents an integration constant. Analytical solutions are not feasible for this expression, given its highly non-linear and composite structure. As a result, we evaluate the integral numerically and present the corresponding evolution of black hole mass in Figure \ref{fig:f10}.

From Figure \ref{fig:f10}, we observe that all mass evolution curves begin at $M=1$ at present time $(z=0)$ and decrease as the redshift increases, consistent with the notion that black holes grow over time due to accretion. For small values of the accretion parameter, such as $A=0.001$, the black hole mass remains nearly negligible across the cosmic timeline. For higher values, the mass increases progressively: at $A=0.002$, the mass reaches about $M\approx0.12$ at early epochs, for $A=0.003$, it approaches $M\approx0.17$ and for $A=0.009$, it can rise as high as $M\approx0.29$ in the distant past. As redshift approaches zero, the increasing trend in black hole mass indicates that black holes are accumulating matter from the surrounding cosmic environment. This behavior implies that scalar field dark energy plays a notable role in the accretion process at later cosmic times. Conversely, the slower growth rate observed at higher redshifts suggests that, during the early Universe, black holes had restricted interaction with scalar field energy, and substantial accretion has only become prominent in more recent epochs \cite{Wang23,Lima2010,Barrow1988}.

Analyzing black hole accretion in the framework of $f(Q,L_{m})$ gravity provides a window into the interaction between spacetime geometry and non-minimally coupled matter. Black holes can be used to study modified gravity effects and hold a record of the Universe's evolution, encoded in their mass history. The scalar field, influenced by the non-metricity $Q$ and matter Lagrangian $L_{m}$, modifies the energy conditions around the black hole and therefore impacts the accretion rate. This establishes a new observationally relevant link between large-scale cosmic acceleration and compact astrophysical systems. 
\section{Conclusion}\label{sec10}
\hspace{0.5cm} This work examines how a scalar field characterized by the energy density expression $\rho_{\phi}=\rho_{c0}tanh(A+Bz)$ influences cosmic evolution within the modified gravity framework $f(Q,L_{m})$ gravity. By employing a linear form of the function, $f(Q,L_{m})=\beta Q+\delta L_{m}$, we focus on the effects of simple linear interactions between the geometric and matter sectors. From the modified field equations, we derived the Hubble parameter analytically and constrained the model parameters using a combined MCMC analysis with $31$ CC, $15$ BAO, DESI DR2 BAO and $1701$ Pantheon+ supernova data points. The best-fit values obtained are $H_{0}=74.284^{+4.155}_{-4.275}$, $\Omega_{m0}=0.326^{+0.093}_{-0.072}$ and $B=-0.001^{+0.030}_{-0.030}$, which are in strong agreement with recent observational bounds.

We analyzed the cosmological evolution through various diagnostic tools. The deceleration parameter $q(z)$ starts at $q\approx0.5$, transitions to acceleration at redshift $z_{tr}=0.5914$ and asymptotically approaches $q=-1$ in the future, mimicking a de Sitter phase. The present value $q_{0}=-0.5167$ is observationally viable. The energy density and pressure profiles confirm that while matter dominates at early times, it decays in the future, and the scalar field increasingly dominates, supporting late-time acceleration. The scalar field’s EoS parameter $\omega_{\phi}$ evolves from $-0.9948$ to $-1$ and the present value $\omega_{0}=-0.99973$ reaffirms its near–$\Lambda$CDM behavior.

The evolution of the kinetic term $\dot{\phi}^{2}$ and scalar potential $V(\phi)$ is physically consistent, remaining positive throughout and indicating smooth dark energy evolution. The density parameters $\Omega_{m}$ and $\Omega_{\phi}$ reveal a successful transition from matter dominance to scalar field dominance, with present values $\Omega_{m0}=0.3201$ and $\Omega_{\phi0}=0.672$, which matches observational data well.

Furthermore, the model predicts a present cosmic age of $t_{0}=13.51$ Gyr, derived from the condition $H_{0}t_{0}=1.00382$, which agrees with Planck and other cosmic age estimates. The behavior of the statefinder parameters shows a Chaplygin gas–like evolution $(r>1, s < 0)$ in the early Universe, converging smoothly to $\{r, s\} = \{1, 0\}$ which is consistent with $\Lambda$CDM at late times.

Lastly, we explored black hole mass evolution under scalar field accretion using a generalized Babichev-type formalism. The mass $M(z)$ decreases with redshift, with higher accretion constants leading to more substantial mass growth. This behavior reinforces the model's ability to describe matter-energy transfer in a dynamically evolving Universe.

Overall, our findings demonstrate that the proposed $f(Q,L_{m})$ scalar field model accurately describes the full cosmic history—from early deceleration to late-time acceleration—while remaining consistent with a broad range of current cosmological observations. The model’s predictive strength and observational agreement highlight its potential as a compelling alternative to standard $\Lambda$CDM cosmology.

\textbf{DATA AVAILABILITY STATEMENT}: There are no new data associated with this article.\\

\end{document}